# How Well Do We (and Will We) Know Solar Neutrino Fluxes and Oscillation Parameters?

J.N. Bahcall and P.I. Krastev

*School of Natural Sciences, Institute for Advanced Study*

*Princeton, NJ 08540*

## Abstract

Individual neutrino fluxes are not well determined by the four operating solar neutrino experiments. Assuming neutrino oscillations occur, the pp electron neutrino flux is uncertain by a factor of two, the $^8$B flux by a factor of five, and the $^7$Be flux by a factor of forty-five. For matter-enhanced oscillation (MSW) solutions, the range of allowed differences of squared neutrino masses, $\Delta m^2$, varies between $4 \times 10^{-6}$ eV$^2$ and $1 \times 10^{-4}$ eV$^2$, while $4 \times 10^{-3} \leq \sin^2 2\theta \leq 1.5 \times 10^{-2}$ or $0.5 \leq \sin^2 2\theta \leq 0.9$. For vacuum oscillations, $\Delta m^2$ varies between $5 \times 10^{-11}$ eV$^2$ to $1 \times 10^{-10}$ eV$^2$, while $0.7 \leq \sin^2 2\theta \leq 1.0$. The inferred ranges of neutrino parameters depend only weakly on which standard solar model is used. Calculations of the expected results of future solar neutrino experiments (SuperKamiokande, SNO, BOREXINO, ICARUS, HELLAZ, and HERON) are used to illustrate the extent to which these experiments will restrict the range of the allowed neutrino mixing parameters. For example, the double ratio (observed ratio divided by standard model ratio) of neutral current to charged current event rates to be measured in the SNO experiment varies, at 95% confidence limit, over the range: 1.0 (no oscillations into active neutrinos), $3.1^{+1.8}_{-1.3}$ (small mixing angle MSW), $4.4^{+2.0}_{-1.4}$ (large mixing angle MSW), and $5.2^{+5.6}_{-2.9}$ (vacuum oscillations).

We present an improved formulation of the "luminosity constraint" and show that at 95% confidence limit this constraint establishes the best available limits on the rate of creation of pp neutrinos in the solar interior and provides the best upper limit to the $^7$Be neutrino flux. The actual rate of creation of solar neutrinos in the solar interior to the rate predicted by the standard solar model can vary (while holding the CNO neutrino flux constant) between 0.55 to 1.08 for pp neutrinos and between 0.0 and 6.35 for $^7$Be neutrinos.

## I. INTRODUCTION

The purpose of this paper is to determine how well the four pioneering solar neutrino experiments determine the neutrino fluxes and possible mass and mixing parameters. We also explore to what extent the solar neutrino experiments under construction are likely to improve this knowledge.

The reader who wants to quickly digest the main points of this paper is urged to turn immediately to the summary given in the concluding section, Section IX. Since this paper contains many detailed results, even the expert may find it useful to read this conclusion section first in order to obtain an overview of the principal results before becoming involved in the details.

Table I summarizes the latest results for the four pioneering experiments on which the theoretical inferences are based. The rates of the three radiochemical experiments (Homestake [1], GALLEX [2], and SAGE [3]) are given in solar neutrino units (1 SNU = $10^{-36}$ events per target atom per second) and the rate of the water-Čerenkov experiment (Kamiokande, [4]) has been presented as the measured $^8$B neutrino flux in units of cm$^{-2}$s$^{-1}$. We also show in Table I the ratios of the observed to the predicted rates, taking the predicted rates from the recent solar model of Bahcall and Pinsonneault [5] (which includes metal and helium diffusion) and assuming nothing happens to the neutrinos after they are created in the center of the sun (which is implied by the standard electroweak model [6]).

The neutrino fluxes and mixing parameters determined in this paper are all consistent at the 95% confidence level with the four operating experiments.

The four operating solar neutrino experiments do not provide enough information to determine uniquely the solar neutrino spectrum at earth. For example, three of the four experiments (chlorine and the two gallium experiments, GALLEX and SAGE) are radiochemical measurements that determine the number of events above a threshold energy (which is different for each experiment). They cannot distinguish, for example, between a large number of low energy neutrinos and a smaller number of higher energy neutrinos.

One must adopt some theoretical framework in which to answer the question: How well do we know the neutrino fluxes at earth from the individual solar neutrino reactions? We calculate the answer to this question by assuming Mikheyev-Smirnov-Wolfenstein (MSW) [7] or vacuum [8] neutrino oscillations and by adopting the basic picture of the solar neutrino spectrum [9]. We do not discuss sterile neutrinos, which would increase the set of possible solutions [10].

After more than thirty years of investigation, there is essentially universal agreement on the validity of the basic picture of neutrino production in the sun [9], namely, that the principal neutrino sources are pp, pep, $^7$Be, $^8$B, $^{13}$N, and $^{15}$O, each source with a characteristic, known energy spectrum. This general picture has become so accepted that it is sometimes referred to as "model-independent"(see, e. g., [11,12]). There is also widespread agreement among workers in the field that the standard solar model gives a reasonable quantitative estimate of the neutrino fluxes from each source. Individual solar models differ somewhat in their predicted fluxes, but all recently-published models give essentially the same neutrino fluxes when the same input parameters are used [13,14]. Moreover, as we shall see from the detailed calculations presented in this paper, the dispersion in predicted neutrino fluxes between the published solar models is very small compared to the range of allowed fluxes permitted by the existing experiments. The currently-dominant uncertainties are experimental not theoretical.

In order to obtain the smallest (plausible) estimate of the uncertainties in the neutrino fluxes at earth, we adopt the predictions of the standard solar model and the estimated uncertainties in those predictions. Table II gives the relevant predictions of the standard solar model of Bahcall and Pinsonneault [5], which includes helium and heavy element diffusion. The first column of Table II identifies the neutrino source, the second column gives the neutrino energy range, and the third column gives the predicted fluxes and their associated ($1\sigma$) uncertainties.

We present in Section II the best-fit MSW [7] and vacuum neutrino oscillation [8] solutions to the most recent published data from solar neutrino experiments (see Table I) and the most detailed solar model [5]. In Section III, we show that the total range of allowed neutrino oscillation solutions is very large even if we require agreement with the standard solar model. The best-fit solutions and the total allowed range depend only weakly upon whether we adopt a 1988, 1992, or a 1995 version of the standard solar model.

A number of authors [15–20] have discussed previously a constraint on a linear combination of the solar neutrino fluxes that is required by the assumption that the Sun is currently supplying energy by nuclear fusion reactions in its interior at a rate that is essentially equal to the observed photon luminosity. We present in Section IV this luminosity constraint in a more complete and precise form.

We calculate the allowed rates of creation for solar neutrinos in Section V (for MSW solutions) and in Section VI (for vacuum oscillations). We use the luminosity constraint to limit the allowed rates. Our approach in these two sections is somewhat similar to the approach adopted by Hata and Langacker [12], who however do not consider vacuum oscillations or the 1995 standard solar model and used a less complete statement of the luminosity condition. On the other hand, Hata and Langacker [12] provide powerful arguments that non-standard solar models are not a viable solution to the solar neutrino problem, a topic not discussed in the present paper.

For future solar neutrino experiments ($^8$B: SuperKamiokande [21], SNO [22], ICARUS [23]; $^7$Be: BOREXINO [24]; and pp: HELLAZ [25] and HERON [26]), we calculate in Section VII the best estimates of the expected event rates using the best-fit MSW and vacuum oscillation solutions (given in Section II) to the four existing experiments. We also calculate the expected results of these experiments to study how they can further constrain the allowed regions in the space of neutrino oscillation parameters.

We calculate in Section VIII the expected ratio of neutral current to charged current events in the SNO detector. We present results assuming that no oscillations occur or that either MSW or vacuum oscillations occur. As was pointed out in the original proposal for a SNO detector [22], this ratio is relatively independent of solar model considerations and can be used to discriminate between broad classes of hypothetical solutions to the solar neutrino problems.

## II. BEST-FIT SOLUTIONS

In this section, we present best-fit MSW and vacuum oscillation solutions to the most recent published data from the solar neutrino experiments (see Table I) and the most detailed solar model (see Table II). We also compare the allowed regions for three different solar models, which we will sometimes refer to as the 1988 [13], 1992 [27], and 1995 [5] standard solar models. The main physical difference between the 1995 model [5] and the two earlier models is the inclusion of both heavy element and helium element diffusion in the calculations, which leads to higher predicted event rates in all three detectors. The other changes in the 1995 model result from refinements of input data.

Many authors have reported the results of refined studies of the MSW [12,20,28–33] and the vacuum [30,31,33,34] solutions of the solar neutrino problems. The techniques for this analysis are therefore well documented in the literature and we only note briefly here those aspects of the calculation that require special mention. Our $\chi^2$ analysis of the data follows closely the prescription of Fogli and Lisi [20], except we treat the GALLEX and SAGE results as separate measurements. We have verified that the allowed regions for the parameters $\Delta m^2$ and $\sin^2 2\theta$ do not differ significantly from the ones obtained by combining the data from the two gallium experiments into a single data point. We include the published energy resolution and trigger efficiency of the Kamiokande detector [35]. We do not include a day-night comparison in the rates of the Kamiokande detector, since these data are not yet available for the latest published average event rate [4]. This omission is not important since previous studies [29,35] have shown that the nonobservation of the day-night effect only excludes a region in oscillation parameter space that is otherwise excluded by combining the results from the four operating experiments. We therefore do not include in the calculations described here the theoretical effect of neutrino regeneration in the earth (see [36]). Based upon exploratory calculations we have done, we expect that the inclusion of the available day-night data and the inclusion of the calculated earth regeneration effect, would only slightly deform the large mixing allowed region calculated in this paper and affect only marginally the rest of our results. For the MSW solution, we use the analytical description of the neutrino survival probabilities from [37] which allows the averaging over the neutrino production regions and the neutrino spectra to be done accurately with a reasonable amount of computer time. We use the neutrino interaction cross sections for each detector given in [13].

The MSW mechanism provides a good fit to the data. There are two local minima of the $\chi^2$ function in the $\Delta m^2 - \sin^2 2\theta$ plane. The allowed region around the deepest minimum, which has the very small value of $\chi^2_{\min} = 0.31$, occurs at

$$\Delta m^2 = 5.4 \times 10^{-6} \text{ eV}^2, \tag{1a}$$

$$\sin^2 2\theta = 7.9 \times 10^{-3}. \tag{1b}$$

The minimum defined by Eq. (1) is usually referred to as the "small mixing angle solution." In contrast, the "large mixing angle solution" has $\chi^2_{\min} = 2.5$, which is relatively large but acceptable for two degrees of freedom. The large mixing angle solution occurs at

$$\Delta m^2 = 1.7 \times 10^{-5} \text{ eV}^2, \tag{2a}$$

$$\sin^2 2\theta = 0.69. \tag{2b}$$

Figure 1 shows, assuming MSW oscillations occur, the allowed regions for the three different solar models (1988, 1992, 1995), together with the points of the local $\chi^2$ minima. The precise positions of the allowed regions and their shape depend not only on the measured rates in the solar neutrino experiments, but also somewhat on the solar model. The higher rates predicted in the model [5] require a stronger depletion of the electron neutrino flux from the Sun in order to account for the experimental data. Thus the parameters $\Delta m^2$ and $\sin^2 2\theta$ which minimize $\chi^2$ change as shown in the figure. Both the small and the large mixing angle allowed regions shift slightly toward the center of the figure, where the survival probability is lowest and the flux suppression strongest.

Vacuum neutrino oscillations provide a somewhat worse but still acceptable fit to the data. The minimum in $\chi^2$ is $\chi^2_{\min} = 2.5$ and occurs at

$$\Delta m^2 = 6.0 \times 10^{-11} \text{ eV}^2, \tag{3a}$$

$$\sin^2 2\theta = 0.96. \tag{3b}$$

There are several disconnected allowed regions in which the local minima of $\chi^2$ are larger than the global minimum whose location is specified by Eq. (3).

Figure 2 shows, assuming vacuum neutrino oscillations occur, the allowed regions that were determined using the three different standard solar models (1988, 1992, and 1995). The inferred neutrino parameters are relatively insensitive to the many successive improvements in the standard solar models. The higher predicted event rates for the 1995 solar model require higher suppression of the electron neutrino flux, which causes the small mixing angle allowed region to shift to larger values of $\sin^2 2\theta$ and the large mixing angle allowed region to shift to smaller values of $\sin^2 2\theta$.

### III. SURVIVAL PROBABILITIES

In this section, we determine the extreme range of the allowed survival probabilities for each neutrino flux. We use the term "survival probability" to mean the probability that a neutrino created in the solar interior will remain an electron type neutrino until it reaches the detector on earth. All of the survival probabilities calculated in this paper are averaged over the solar interior using the 1995 detailed solar model [5].

We define both "detector-dependent survival probabilities" (which have previously been used in the literature) and "detector-independent survival probabilities" (not previously used) and present results for both sets of quantities. We calculate the maximum and minimum fluxes consistent with the available data for both the MSW effect and for vacuum oscillations. The results given here permit the calculation of what may be observed in future experiments.

We assume that the published event rates and experimental errors (see Table I) are correct for the four operating experiments. We also assume that the calculated standard model production rates for neutrinos (see Table II) are correct within their published uncertainties. With these assumptions, we determine the allowed ranges of survival probabilities for both MSW and for vacuum neutrino oscillations. From our experience, we infer that authors are more likely to underestimate rather than overestimate uncertainties when reporting either experimental or theoretical results. Since we adopt the published uncertainties, the allowed range we find for the survival probability of each neutrino flux is likely to be a lower limit to the actual permitted range.

### A. Methods of Calculation

For a dense set of representative values of $\Delta m^2$ and $\sin^2 2\theta$ within the 95% confidence limits indicated by Fig. 1 and Fig. 2, we calculate the survival probability, $P(\nu_e \to \nu_e, E)$, for electron-type neutrinos. For continuum neutrino sources, we average the survival probability over neutrino energy using the known $\nu_e$ energy spectrum when created in the sun. We carry out the calculations with, or without, the additional weight of the energy-dependent neutrino interaction cross section for the detector of interest. For each $\Delta m^2$ and $\sin^2 2\theta$ chosen within the allowed region, we compute the partial contribution of each flux to the signal in each detector by using the neutrino survival probabilities calculated as described earlier. We find the minimum and maximum values for each survival probability by searching among the allowed set of solutions.

The reader may be surprised that we present in this section the limiting contributions of different neutrino sources to experiments that have not yet been performed. However, the logic we use in these calculations is the same as we use for the operating experiments. We search through the set of solutions consistent with the already available data and find those solutions that maximize or minimize the contributions of a particular neutrino source to each experiment, whether or not the experiment has been performed. For future experiments, we do not of course have the benefit of direct constraints based upon the counting rate in that detector.

### B. Detector-Dependent Survival Probabilities

In this subsection, we calculate survival probabilities in the way that they are usually determined in the literature, namely, with a weight that is proportional to the interaction cross section in a specified detector. In the following subsection, we calculate survival probabilities that are independent of the characteristics of any specific detector.

Table III and Table IV present the results of numerical calculations for many different theoretical solutions based upon either the MSW effect [7] or upon vacuum neutrino oscillations [8]. Each row gives, for a specific solar neutrino experiment, the minimum and maximum allowed contribution relative to the standard model prediction of the pp, $^7$Be, and $^8$B neutrino sources. The calculations apply to the operating experiments (chlorine, gallium, and Kamiokande) as well as to planned experiments (SuperKamiokande, SNO, BOREXINO, ICARUS, HELLAZ, and HERON).

For each experiment, we present in Table III the extreme MSW survival probabilities weighted by the cross sections for interactions of electron-type neutrinos. The weighted survival probabilities for neutrinos of source $i$ are defined by the following equation:

$$(\text{Weighted Survival Probability})_i = \frac{\int dE \ \text{Flux}_i(E) \times \sigma(E) \times P_i(E)}{\int dE \ \text{Flux}_i(E) \times \sigma(E)}, \qquad (4)$$

where $\sigma(E)$ is the interaction cross section and $P_i(E)$ is the survival probability for electron-type neutrinos averaged over the production region of the $i$-th neutrino source. The integrals extend over the range in which the neutrino energy spectrum, $\text{Flux}_i(E)$, is non-zero. The weighted survival probabilities defined in Eq. (4) are the ratios of the actual event rates to the standard model predicted rates (standard electroweak plus standard solar model). For brevity, we use the notation $R_i \equiv$ (weighted survival probability)$_i$ in Table III and Table IV (also in Table V). Thus the first entry (0.0045) under R($^7$Be)$_{\min}$ in the first row of Table III is, for the MSW solution, the largest fractional reduction (relative to the standard solar model) of the $^7$Be contribution to the chlorine experiment that is consistent at 95% C.L. with all the experiments. For example, to obtain the maximum $^8$B contribution to the chlorine experiment in SNU, one has to multiply the entry in Table III, 0.56, times the standard model prediction of 7.36 SNU [5] to get 4.12 SNU. This large value, which considerably exceeds the experimental result given in the first row of Table I, is nevertheless allowed at 95% C.L. by the $\chi^2$ analysis because of the large theoretical uncertainty (14% at $1\sigma$) of the $^8$B neutrino flux in the standard solar model. This uncertainty amounts to 1 SNU of the event rate in the chlorine detector.

Table III shows that, for MSW oscillations, the pp contribution to the GALLEX and SAGE experiments is constrained to vary by at most a factor of two. The pp flux is the best determined of the neutrino fluxes. In the HELLAZ and HERON experiments, the expected range of the pp rate is less than a factor of two since these neutrino-electron scattering experiments are sensitive to both charged and neutral currents.

The weighted average $^8$B flux that is measured in the Kamiokande experiment is also reasonably well determined (slightly more than a factor of two uncertainty), but the $^8$B contributions to the radiochemical experiments (chlorine and gallium experiments) is less well constrained (more than a factor of three uncertainty). The radiochemical experiments have much lower energy thresholds than the Kamiokande experiment, which is also somewhat sensitive to the neutral current contribution.

The $^7$Be contribution to the radiochemical experiments (chlorine and gallium) can vary between approximately 1% and 65% of the contribution predicted by the standard solar model. The rate for BOREXINO is predicted to lie within 22% to 72% of the standard rate. This variation is much less than is allowed in the gallium or chlorine experiments because BOREXINO is sensitive to both charged and neutral currents. The lower limit for BOREXINO corresponds to essentially all but about 1% of the electron neutrinos transforming to muon or tau neutrinos.

Table IV is identical to Table III except that the weighted survival probabilities are computed for the mechanism of vacuum neutrino oscillations. The survival probabilities have been averaged, as described in ref. [30], over the distance between the Sun and the Earth, which changes during the year due to the eccentricity of the Earth's orbit. About 128 points (independent computations) for different positions on the Earth's orbit are needed to compute the average survival probability with an accuracy of 0.1%.

For vacuum oscillation solutions, the range of allowed solutions is smaller than for the MSW effect. In fact, the pp neutrino flux is well determined. The pp contribution to the gallium experimental rate is 58% ± 9% of the standard model rate, which corresponds to a fractional rate of 70% ± 7% in the HELLAZ and HERON detectors. The $^8$B contribution is also reasonably well determined, 39%±14%, for the Kamiokande experiment. The predicted rate relative to the standard solar model of the SNO and ICARUS experiments shows a larger variation since in this section we are considering only charged-current reactions for these two detectors.

Perhaps the most surprising result for the vacuum neutrino oscillations is the fact that the $^7$Be contribution could be essentially equal to the standard model prediction. This result shows that some previous analyses may have oversimplified the situation when they have concluded that the solar neutrino problems can be summarized by the fact that $^7$Be neutrinos are missing.

The extreme values for the masses and mixing angles do not always occur for extreme values of the neutrino fluxes. The extreme values of the neutrino parameters can be determined directly from the data files used to plot Figure 1 and Figure 2. For MSW solutions, the range of allowed $\Delta m^2$ varies between $4 \times 10^{-6}$ eV$^2$ and $1 \times 10^{-4}$ eV$^2$, while $4 \times 10^{-3} \leq \sin^2 2\theta \leq 1.5 \times 10^{-2}$ or $0.5 \leq \sin^2 2\theta \leq 0.9$. For vacuum oscillations, the $\Delta m^2$ varies between $5 \times 10^{-11}$ eV$^2$ to $1 \times 10^{-10}$ eV$^2$, while $0.7 \leq \sin^2 2\theta \leq 1.0$.

## C. Detector-Independent Survival Probabilities

Are there detector-independent bounds that one can place on the survival probabilities? The bounds given in Table III and Table IV depend manifestly upon the interaction cross sections of each detector. These detector-dependent limits are relevant when thinking about what has been learned from each experiment. However, one needs to consider detector-independent bounds if one wants to determine the allowed regions permitted by all the experiments taken together.

We have calculated unweighted (detector-independent) survival probabilities by searching in the complete parameter space of the two-component MSW and vacuum neutrino oscillations for solutions that are consistent at the 95% confidence level with the data summarized in Table I for the four ongoing experiments. The detector-independent average survival probabilities are defined as follows:

$$(\text{Average Survival Probability})_i = \frac{\int dE \ \text{Flux}_i(E) \times P_i(E)}{\int dE \ \text{Flux}_i(E)}. \tag{5}$$

Table V summarizes the extreme values for the detector independent average survival probabilities for electron-type neutrino fluxes that are consistent with either MSW or vacuum oscillations. The range of allowed survival probabilities is a factor of two for pp neutrinos, a factor of more than five for $^8$B neutrinos, and a factor of more than forty for the $^7$Be neutrinos.

## IV. THE LUMINOSITY CONSTRAINT

Each time a neutrino from a specified neutrino-producing reaction is created in the sun a fixed amount of usable (thermal) energy is supplied to the interior of the star. Since some of the nuclear reactions that are responsible for solar energy generation also produce neutrinos, the solar luminosity, $L_\odot$, can be written as a linear combination of solar neutrino fluxes [15–20]. The luminosity constraint on the neutrino fluxes takes the form (cf. Eq. 3.36 of [9])

$$\frac{L_\odot}{4\pi r^2} = \sum_j (Q - \langle E \rangle)_j \, \phi_j \,, \tag{6}$$

where $r$ is 1 A.U. (1.496 × 10$^{13}$ cm), $Q$ is the energy released by the associated fusion reactions, $\langle E \rangle$ is the average energy loss by neutrinos, and $\phi_j$ ($j$ = pp, $^7$Be, $^8$B, etc.) is the neutrino flux at earth (if nothing happens to the neutrinos after they are created). The average energy loss by the neutrinos has been calculated in Appendix A of [13], which takes account of small corrections for the thermal motion of the interacting particles.

The explicit form of the luminosity constraint is

$$\frac{L_\odot}{4\pi r^2} = \sum_j \alpha_j \phi_j \;, \tag{7}$$

where the eight coefficients, $\alpha_j$, are given in Table VI and the solar constant $(L_\odot/4\pi r^2)$ is [5] 1367 Wm$^{-2}$. The numerical values given in Table VI include the small corrections due to the thermal motion of the solar particles [13]. For numerical applications, it is convenient to rewrite Eq. (7) in the following dimensionless form:

$$1 = \sum_i \left(\frac{\alpha_i}{10 \text{ MeV}}\right)\left(\frac{\phi_i}{8.532 \times 10^{10} \text{ cm}^{-2}\text{s}^{-1}}\right) \tag{8}$$

In applications, the linear relation given in Eq. (7) must be supplemented by the additional constraint:

$$\phi\left(^7\text{Be}\right) + \phi\left(^8\text{B}\right) \;\leq\; \phi(\text{pp}) + \phi(\text{pep}) \;. \tag{9}$$

The physical basis of Eq. (9) is that the $^3$He nuclei, which ultimately give rise to $^7$Be and $^8$B neutrinos via the nuclear reaction $^3$He$(\alpha,\gamma)^7$Be, are created by pp and pep reactions. One pp or pep reaction must occur in order to supply the $^3$He nucleus that is burned each time a $^7$Be or $^8$B neutrino is produced. So far as we can tell from the published literature, Eq. (9) has not been implemented in previous applications of the luminosity constraint to the solar neutrino problem. In principle, Eq. (7) considered by itself permits a $^7$Be neutrino flux that is twice as large as is allowed by Eq. (9). Since the $^{14}$N$(p,\gamma)^{15}$O reaction is the slowest process in the CNO cycle, one must also have

$$\phi\left(^{15}\text{O}\right) \;\leq\; \phi\left(^{13}\text{N}\right) \;. \tag{10}$$

The luminosity constraint sets upper bounds on the allowed neutrino fluxes. Since somewhat different amounts of energy release are associated with each of the neutrino fluxes, the upper bounds depend somewhat upon the neutrino branch being considered. We find from Eqs. (7)–(10) the following upper limits for the neutrino fluxes:

$$\phi(\text{pp}) \;\leq\; 6.51 \times 10^{10} \text{ cm}^{-2}\text{s}^{-1} \;, \tag{11a}$$

$$\phi(\text{pep}) \;\leq\; 7.16 \times 10^{10} \text{ cm}^{-2}\text{s}^{-1} \;, \tag{11b}$$

$$\phi(^7\text{Be}) \;\leq\; 3.33 \times 10^{10} \text{ cm}^{-2}\text{s}^{-1} \;, \tag{11c}$$

$$\phi(^8\text{B}) \;\leq\; 4.32 \times 10^{10} \text{ cm}^{-2}\text{s}^{-1} \;, \tag{11d}$$

$$\phi(\text{CNO}) \leq 3.41 \times 10^{10} \text{ cm}^{-2}\text{s}^{-1} .\tag{11e}$$

The upper limits on the fluxes of $^7$Be and $^8$B neutrinos are achieved [see Eq. (9)] when $\phi(\text{pp}) = \phi(^7\text{Be})$ or $\phi(\text{pp}) = \phi(^8\text{B})$. In deriving Eq. (11e), we have assumed that $\phi(^{13}\text{N}) \simeq \phi(^{15}\text{O})$ when the CNO cycle is dominant.

What are the principal assumptions that are required to derive Eqs. (7)–(10)? The fundamental assumption is that nuclear fusion among light elements currently generates energy in the solar interior at a rate equal to the measured solar photon luminosity. Gravitational energy generation is neglected; this causes an error in Eq. (7) of only [13] $-0.03\%\ L_\odot$ (as estimated from the standard solar model). The abundance of $^3$He nuclei is also assumed to be in equilibrium, which is not strictly correct. In the outer regions of the solar core, $^3$He is continually produced (ultimately by the pp and pep reactions), but the temperature is too low to burn $^3$He at the equilibrium rate [via the reactions $^3\text{He}(^3\text{He},2p)^4\text{He}$ and $^3\text{He}(\alpha,\gamma)^7\text{Be}$]. We have calculated the departure $\delta$ from $^3$He equilibrium in the 1995 standard solar model [5] and find

$$\delta \equiv \left[\phi(\text{pp}) + \phi(\text{pep}) - 2\phi(^3\text{He}-^3\text{He}) - \phi\left(^7\text{Be}\right) - \phi\left(^8\text{B}\right)\right] / \left[\phi(\text{pp}) + \phi(\text{pep})\right] = -0.04\% ,\tag{12}$$

where we have introduced for notational convenience a "fictional neutrino flux," $\phi(^3\text{He}-^3\text{He})$, produced by the $^3\text{He}-^3\text{He}$ reaction. Here $\phi(^3\text{He}-^3\text{He})$ is the flux that would be produced if each $^3\text{He}-^3\text{He}$ reaction produced a neutrino. [The export nuclear energy generation subroutine available from one of us (JNB) calculates this fictional flux together with the real neutrino fluxes.]

We conclude that Eqs. (7)–(10) are, when taken together, a statement of the luminosity constraint that is accurate to better than 1% for solar models in which the current nuclear energy generation equals the observed solar luminosity.

## V. ONE NONSTANDARD SOLAR NEUTRINO FLUX: MSW SOLUTIONS

In this section and in the following section, we determine the extent to which existing experiments and the luminosity constraint together bound the allowed values of the neutrino fluxes at earth. We relax the assumption that all of the calculated standard model fluxes are correct within their quoted uncertainties. We permit one of the major fluxes (either pp, $^7$Be or $^8$B) to vary as a free parameter, together with $\Delta m^2$ and $\sin^2 2\theta$. In the present section, we derive the allowed ranges for the principal neutrino fluxes assuming the correctness of the MSW solution [7] of the solar neutrino problem and in the following section we derive allowed ranges for the solutions involving vacuum oscillations [8].

In Section III, we analyzed the range of allowed neutrino fluxes observed at earth, while requiring consistency with the standard solar model calculations. In the present section and in the following section, we study the rate of creation of neutrinos in the solar interior, while only requiring consistency with the luminosity constraint, Eqs. (7)–(10). When calculating the allowed range of the $^8$B neutrinos, we keep the other neutrino fluxes fixed at their standard model values (since the $^8$B flux does not enter significantly the luminosity constraint). While varying either the pp (or the $^7$Be) neutrino flux, we adjust the $^7$Be (or the pp) flux at the value required to satisfy the luminosity constraint (with the minor fluxes fixed at their standard solar model values.) We shall see that the empirical limits on the creation rates of solar neutrinos are much larger than the quoted theoretical uncertainties, so it is not really important which precise values we choose for the fluxes not being systematically varied (as long as we require consistency with the luminosity constraint).

We define fitting factors, $f_i$, which are the ratios of the fluxes actually created in the center of the Sun to the fluxes predicted by the standard (1995) solar model. Thus:

$$f_i \equiv \frac{\phi(i)}{\phi_{\text{SSM}}(i)} \ . \tag{13}$$

Using the three qualitatively different experiments, we can determine three parameters, $\Delta m^2$, $\sin^2 2\theta$ and one of the fluxes; we keep the other fluxes fixed at their standard model values (within the published theoretical uncertainties). We have varied the $f_i$ over a broad range compatible with the luminosity constraint, Eqs. (7)–(10). For each value of $f_i$, we carry out the minimization of $\chi^2$ over $\Delta m^2$ and $\sin^2 2\theta$ within the region for which significant transitions between different flavors occur: $10^{-4} \leq \sin^2 2\theta \leq 1.0$, $10^{-9}$ eV$^2 \leq \Delta m^2 \leq 10^{-3}$ eV$^2$. The positions of the $\chi^2$ minima depend on the values of the $f_i$. In fact, as shown in ref. [38], if $f(^8\text{B}) \neq 1$, the 95% C.L. regions for a particular model can shift significantly with respect to those shown in Fig. 1.

Figure 3 shows the results for the three neutrino ratios, $f_i$. Each $\chi^2$ that is plotted is a function of three independent parameters, $\Delta m^2$, $\sin^2 2\theta$, and one of the major neutrino fluxes (pp, $^7$Be or $^8$B). We display only the minimum (in $\Delta m^2$ and $\sin^2 2\theta$) $\chi^2$ as a function of $f_i$. The luminosity constraint requires that each flux lie within the vertical lines in Fig. 3. The horizontal lines mark the 68% confidence limit ($\chi^2 = \chi^2_{\min} + 1.0$).

Figure 3 shows that the bounds imposed by the luminosity constraint and the 95% confidence limits on the fluxes are:

$$0.55 \leq f(\text{pp}) \leq 1.08, \tag{14a}$$

$$0.0 \leq f(^7\text{Be}) \leq 6.35, \tag{14b}$$

$$0.37 \leq f(^8\text{B}) \leq 2.44. \tag{14c}$$

The pp neutrino flux is the best determined flux. Both the upper and lower limits for the pp neutrinos are established by the luminosity constraint. The flux of pp neutrinos could at most vary by a factor of two and be consistent with the luminosity constraint. The flux of $^7$Be neutrinos is the least well determined. All of the data are consistent with a vanishingly small $^7$Be neutrino flux. The upper limit to the $^7$Be neutrino flux is fixed by the luminosity constraint.

The Kamiokande and the chlorine data constrain the $^8$B neutrino flux to lie within a factor of three of the standard solar model prediction. The $^8$B neutrino flux is practically unconstrained by the solar luminosity since proton capture reaction by $^7$Be nuclei plays only a very minor role in the production of solar energy if the standard solar model is even approximately correct. [In fact, the contribution of $^8$B neutrinos to the right hand side of Eq. (7) is only about $10^{-4}$ of the contribution of the pp neutrinos in the standard solar model.] The two minima in $\chi^2$ in Fig. 3c correspond to the small ($\sin^2 2\theta < 0.1$; minimum at $f(^8\text{B}) = 0.96$) and the large ($\sin^2 2\theta \geq 0.1$; minimum at $f(^8\text{B}) = 1.6$) mixing-angle solutions; the relative maximum between the two minima corresponds to the transition from the small mixing angle to the large mixing angle solution.

## VI. ONE NONSTANDARD SOLAR NEUTRINO FLUX: VACUUM OSCILLATIONS

In this section, we describe an analysis of the vacuum neutrino oscillation solution that is similar to the MSW analysis discussed in the previous section. We consider $\sin^2 2\theta$ and $\Delta m^2$, together with one of the major fluxes, as free parameters. In each case the sum of the neutrino fluxes satisfies the luminosity constraint, Eqs. (7)–(10). We have varied $\Delta m^2$ and $\sin^2 2\theta$ within the following limits: $0.4 \leq \sin^2 2\theta \leq 1.0$; $10^{-12}$ eV$^2 \leq \Delta m^2 \leq 10^{-9}$ eV$^2$.

Figure 4 shows the allowed range for the pp, $^7$Be and $^8$B fluxes, respectively. The bounds imposed by the luminosity constraint and the 95% confidence limits on the fluxes are:

$$0.55 \leq f(\text{pp}) \leq 1.08, \tag{15a}$$

$$0.0 \leq f(^7\text{Be}) \leq 6.35, \tag{15b}$$

$$0.55 \leq f(^8\text{B}) \leq 2.84. \tag{15c}$$

The pp neutrino flux is (as in the MSW solutions) the best determined flux. Both the upper and the lower limits are established by the luminosity constraint. The minimum $\chi^2$ occurs slightly above $f(\text{pp}) = 1$. There is no lower-limit constraint on the $^7$Be flux, which could be zero consistent with all of the experimental data. The upper limit for the $^7$Be neutrinos is established by the luminosity constraint.

The most ironic result for the $^8$B neutrinos is that the data suggest (at $1\sigma$ significance level) a higher neutrino flux than in the standard solar model ($f(^8\text{B}) \geq 1$) if the vacuum neutrino solution is correct. Nearly all the published research on "non-standard" solar models over the last 25 years has had the stated goal of producing a lower (not a higher) $^8$B neutrino flux than in the standard solar model.

## VII. FUTURE DETECTORS

Two new solar neutrino experiments, SuperKamiokande [21] and SNO [22], are expected to start taking data in 1996-97. The BOREXINO detector [24] is likely to become operational before the end of the century. Recently, a 600 ton module of the ICARUS [23] detector has been approved. Even further in the future, two ambitious helium detectors, HELLAZ [25] and HERON [26], whose purpose is to detect the basic pp neutrinos, are currently being studied as laboratory prototypes. An $^{127}$I detector is under development, but the neutrino absorption cross sections are not well enough known at present to permit a detailed theoretical analysis of the kind carried out in this paper [39].

In this section, we make an initial appraisal of the potential of these future experiments to further constrain the allowed regions in the parameter space of $\Delta m^2$ and $\sin^2 2\theta$. In Section VII A, we calculate the best-estimate rates for all five of the future experiments, assuming the correctness of either the small or the large mixing angle MSW solution or of vacuum oscillations. We use in Section VII A values for the neutrino parameters that minimize $\chi^2$ for the four operating experiments. In Section VII B, we calculate the 'future' allowed regions in neutrino parameter space that will exist if one or two of the new experiments are performed. We focus in this subsection on the three experiments (SuperKamiokande, SNO, and BOREXINO) that are in advanced stages of development. In Section VII B, we allow the values of $\Delta m^2$ and $\sin^2 2\theta$ to range over the region of parameter space permitted, at 95% C.L., by the four operating and the assumed future experiments.

### A. Best-Estimate Future Rates

We calculate in this subsection the rates in future experiments that are implied by the results of the four operating experiments and the different neutrino oscillation scenarios. We adopt the best-fit neutrino oscillation parameters given in Eqs. (1)–Eq. (3), the expected characteristics of the detectors specified in the published proposals, and the neutrino interaction cross sections (except for SNO) given in [9]. For SuperKamiokande, we shift the trigger efficiency function of Kamiokande 1.1 MeV lower in energy so that it becomes 50% at 5 MeV. The neutrino cross sections for SNO are taken from [40]. We defer a discussion of the SNO neutral current experiment to the following section. We do not discuss in this paper the potentially powerful SuperKamiokande measurement of the electron recoil spectrum and the important SNO measurement of the shape of the $\nu_e$ energy spectrum. The potential of both of these measurements to discriminate alternative hypotheses is being investigated, using an improved $^8$B neutrino energy spectrum, in a separate study [41].

Table VII gives the best-estimate calculated event rates relative to the predictions of the standard solar model [5]. We also include the range over which the predictions vary if we require consistency, at 95% C.L., with the four operating experiments. The second column of Table VII gives the result that we expect if small-angle MSW oscillations are occuring, the next column gives the results for large-angle MSW oscillations, and the last column gives the results if vacuum oscillations are the explanation of the solar neutrino problems.

The best-estimate predicted event rates are not very sensitive to the assumed neutrino oscillation scenario. Expressed as percentages of the standard model rates, we find that the best-estimates range over the following values: SuperKamiokande (36% ±5%), SNO (26% ±7%), BOREXINO (46% ±24%), ICARUS (28% ±6%), and HELLAZ/HERON (82% ±15%). However, if we consider the 95% confidence limits, the total range of expected event rates is rather large in all of the experiments (see Table VII).

### B. 'Futuristic' Allowed Regions

Figure 5–Figure 8 display the results of our futuristic simulations of the improvements in determining neutrino parameters that are likely to result from measuring the total rates in the new experiments. We have carried out the combined $\chi^2$ analysis as if the calculated rates for future experiments are indeed the rates that will be measured.

The results shown in Figure 5–Figure 8 assume, for specificity, a 5% $1\sigma$ measurement error in each future experiment. We have also carried out calculations assuming 10% and 1% measurement errors for all the experiments. There is a significant improvement in the ability of the experiments to discriminate between different values of neutrino mixing parameters when the assumed errors are reduced from 10% to 5%, but there is very little additional improvement when the errors are reduced from 5% to 1%.

If the small mixing angle MSW solution is correct, will future experiments eliminate the large mixing angle solution and the vacuum oscillation solution? This question is answered in Fig. 5 and Fig. 6.

We assume in calculating the results that are displayed in Fig. 5 and Fig. 6 that the small angle MSW solution (the minimum $\chi^2$ solution) is the correct description of neutrino propagation. The regions that will be allowed in neutrino parameter space are plotted in Fig. 5a-c and Fig. 6a-c if just one new experiment is performed (SuperKamiokande, SNO, or Borexino). Figure 5a-c show the regions that will be allowed for MSW solutions and Fig. 6a-c show the regions that will be allowed for vacuum oscillation solutions.

We also performed the same type of analysis by adding two "new" experiments, SuperKamiokande and SNO, to the four already existing experiments. Figure 5d and Fig. 6d show the allowed regions when two new experiments are simulated.

Figure 5 shows that 5% measurements of the total event rates will (with our assumption of the correctness of the small-angle MSW solution) essentially rule out at 95% C.L. the large mixing angle solutions. BOREXINO and SNO are particularly effective in this respect. Since the small-angle solution has a very low $\chi^2$ with for the four operating experiments, it is not surprising that one or two additional experiments consistent with this solution will eliminate the large angle solution.

Figure 6 shows the situation is less favorable for the vacuum oscillations. They will be more difficult to rule out if the small mixing angle solution is correct. The one exception is BOREXINO. A measurement of the total flux from the 0.86 MeV $^7$Be line, consistent with the current best-estimate from the MSW solution, would eliminate the currrently-allowed vacuum osciliation solutions (see Fig. 6 c). The minimal $\chi^2$ for the vacuum oscillation region of parameter space, Fig. 6, is relatively large in all four cases: 3.7 for BOREXINO, 4.9 for SuperKamiokande, 4.7 for SNO, and 6.1 for SuperKamiokande plus SNO, but still acceptable at 95% C.L. .

We now consider the opposite question to the one we just answered. If the vacuum mixing solution is correct, will future experiments eliminate the MSW solutions and converge on the true vacuum solution? This question is answered in Fig. 7 and Fig. 8.

We assume in calculating the results that are displayed in Fig. 7 and Fig. 8 that the vacuum oscillation solution with the smallest $\chi^2$ is correct. We see that both BOREXINO and SNO can, with our assumptions, eliminate all, or nearly all, of the allowed regions of the small angle MSW solution. However, the large angle MSW solution will be not eliminated if the best-estimate current vacuum solution is the one that nature has adopted. The vacuum oscillation solutions will be constrained significantly by any one of the three experiments, SuperKamiokande, SNO, or BOREXINO (see Fig. 8). With our assumptions, the combination of SuperKamiokande and SNO will greatly reduce the allowed parameter space for vacuum oscillations.

Larger reductions in the allowed parameter space might occur if, for example, the new experiments yield results that lie at the edge of the parameter space permitted by the four operating experiments.

## VIII. NEUTRAL CURRENT TO CHARGED CURRENT RATIO IN SNO

In this section, we focus on a specific measurement that will be made with the Sudbury Neutrino Observatory [22]. We calculate the expected ratio, (NC/CC), of neutral current events, NC ($\nu_{\text{any}} + {}^2\text{H} \longrightarrow p + n + \nu_{\text{any}}$), to charged current events, CC ($\nu_e + {}^2\text{H} \longrightarrow p + p + e$, only $\nu_e$), in the SNO experiment. Here $\nu_{\text{any}}$ represents the total flux of neutrinos, independent of neutrino flavor. We compute and compare the ratio assuming that there are no neutrino oscillations or that either the hypothesis of MSW or vacuum neutrino oscillations is correct. The best-estimates for the oscillation parameters are again taken from Eqs. (1)–(3), which assume the correctness of the standard model fluxes (within their quoted uncertainties).

In the literature, it has often been stated that the ratio of neutral current to charged current event rates is "independent of solar model predictions" [22], with the implication that the ratio is thereby independent of the total flux (all flavors) of $^8$B neutrinos. This interpretaion of the neutral current to charged current ratio is correct if there is no physics beyond the standard electroweak model that changes the shape of the $^8$B neutrino energy spectrum; solar physics affects the shape of the energy spectrum by at most 1 part in $10^5$ [42]. In the absence of physics beyond the standard electroweak model, the total number of created $^8$B neutrinos cancels out of the ratio since the predicted $^8$B neutrino flux appears in both the numerator and the denominator of the neutral current to charged current ratio. However, if either MSW or vacuum neutrino oscillations occur, the ratio depends somewhat upon the assumed total $^8$B neutrino flux since these solutions change the calculated shape of the neutrino energy spectrum and the best-fit oscillation parameters depend upon the absolute value of the assumed flux. We will explore this dependence in detail in a future publication.

We normalize the neutral current to charged current ratio, (NC/CC), by dividing the observed ratio by the ratio computed assuming the correctness of the standard electroweak model (i. e., no oscillations). This double ratio, DR, is defined to be

$$\left(\frac{\text{NC}}{\text{CC}}\right)_{\text{DR}} \equiv \frac{(\text{NC}/\text{CC})_{\text{observed}}}{(\text{NC}/\text{CC})_{\text{standard}}}. \tag{16}$$

Many of the systematic uncertainties cancel out of the double ratio. Moreover, the double-ratio takes on the simple value of unity if no new physics is occuring.

Table VIII gives the ratio of neutral to charged current event rates for two different assumed energy thresholds (5 MeV and 6 MeV) for the charged current reaction. In each case (no oscillation, MSW, or vacuum oscillations), the upper row assumes a threshold of 5 MeV for the total electron energy and the lower row assumes a 6 MeV threshold. (The threshold of the neutral current reaction is 2.2 MeV.) The values without oscillations of the NC/CC ratio and the double ratio, Eq. (16), are given in the first two rows. In the second and third columns of Table VIII, we give the computed values of the neutral current to charged current ratio, and the double ratio, for $\Delta m^2$ and $\sin^2 2\theta$ that minimize $\chi^2$ [see Eqs. (1) and (3)]. The ranges of the double ratio given in columns four and five have been obtained by varying the oscillation parameters within the 95% C.L. allowed regions (for the four operating experiments), shown in Figs. 1–2, assuming the flux uncertainties of the standard solar model [5]. In computing the charged-current rate, we assumed that the energy of the electron that is produced is equal to the incoming neutrino energy minus the threshold energy (neglecting nucleon recoil and final state interactions as well as the energy resolution of the detector). Lisi (private communication, 1995) has made detailed calculations which show that the assumptions used here are typically accurate to a few percent for MSW oscillations and to about 15 percent for vacuum oscillations.

Table VIII shows that, within relatively large ranges of parameter variations, the measurement of the neutral current to charged current ratio can determine if neutrino oscillations are the correct solution of the solar neutrino problem. For example, a measurement of a double ratio equal to unity with a $1\sigma$ accuracy of 20% would rule out any oscillation solution in Table VIII (i.e., any oscillation solution that does not include sterile neutrinos) at the 99% C.L. However, it may well be difficult to distinguish between different oscillation solutions because of the considerable overlap among the "new physics" solutions in Table VIII.

## IX. CONCLUSIONS

In this paper, we have investigated the range of neutrino oscillation solutions that are allowed by the currently operating solar neutrino experiments and have explored, in a preliminary way, the additional constraints that may be imposed by future experiments. We study the allowed range of neutrino fluxes at earth and the allowed neutrino creation rate at the center of the sun. We summarize in this section some of our principal conclusions.

• The MSW solution provides an excellent description of the results of the four operating experiments. Using the measured values and the quoted uncertainties for the experiments (see Table I) and the solar model predictions and their associated uncertainties (see Table II), the minimum value for $\chi^2$ ($\chi^2_{\min} = 0.31$) occurs at $\Delta m^2 = 5.4 \times 10^{-6}$ eV$^2$ and $\sin^2 2\theta =$ $7.9 \times 10^{-3}$ and is usually referred to as the "small-angle" MSW solution. The "large angle" MSW solution occurs at $\Delta m^2 = 1.7 \times 10^{-5}$ and $\sin^2 2\theta = 0.69$ and is a less good fit to the data ($\chi^2_{\min} = 2.5$).

- Vacuum neutrino oscillations provide an acceptable but not remarkably good fit to the experimental results and the standard model calculations. The minimum in $\chi^2$ ($\chi^2_{\min} = 2.5$) occurs at $\Delta m^2 = 6.0 \times 10^{-11}$ eV$^2$ and $\sin^2 2\theta = 0.96$.

- The best-fit and allowed regions of the oscillation solutions do not depend sensitively upon the assumed solar model. We have used standard solar models published in 1988, 1992, and 1995; the results are rather similar in all cases (see Figure 1 for the MSW solutions and Figure 2 for the vacuum oscillation solutions).

- The first three conclusions in this section (Section IX) are in good agreement with the calculations of previous authors [12,20,29–32,34,33], who made use of earlier results from the solar neutrino experiments and the standard solar models.

- For MSW solutions, the allowed range of the electron neutrino fluxes at earth is large (see Table III). The flux of pp neutrinos is best determined, but even in this case the contribution to the GALLEX and SAGE experiments could vary between 50% to 100% of the predicted standard solar model value. The expected range in the HELLAZ and HERON experiments is between 65% and 100% of the standard model value; the variation is smaller for these future experiments because they have some sensitivity to neutral current reactions. The $^7$Be neutrino flux is poorly constrained. The contribution of $^7$Be to the gallium experiments could range between 1% and 65% of the standard model prediction; the expected range in BOREXINO is between 21% and 72% of the standard rate.

- The range of allowed solutions for vacuum neutrino oscillations is smaller than the allowed range for MSW solutions (see Table IV). The pp contribution to the gallium rate is 58% $\pm$ 9% of the standard model rate, which corresponds to 70% $\pm$ 7% in the HELLAZ and HERON detectors. The $^8$B contribution corresponds to 39% $\pm$ 14% of the standard rate for the Kamiokande experiment. The $^7$Be contribution can vary from approximately 10% to approximately 100% of the predicted rate for the chlorine and gallium experiments, while the expected rate in BOREXINO is between 29% and 98% of the standard model prediction.

- For MSW solutions, the range of allowed neutrino masses is contained between $4 \times 10^{-6}$ eV$^2$ and $1 \times 10^{-4}$ eV$^2$, while $4 \times 10^{-3} \leq \sin^2 2\theta \leq 1.5 \times 10^{-2}$ or $0.5 \leq \sin^2 2\theta \leq 0.9$. For vacuum oscillations, the masses vary between $5 \times 10^{-11}$ eV$^2$ to $1 \times 10^{-10}$ eV$^2$, while $0.7 \leq \sin^2 2\theta \leq 1.0$.

- Detector-independent limits on the allowed ranges of the electron neutrino fluxes can be extracted from the existing experiments (see Eq. 5 and Table V). Considering the combined range of the MSW and vacuum neutrino solutions, the pp flux could be anywhere from 49% to 100% of the predicted standard model value, while the $^8$B flux lies between 11% and 60% of the standard value, and the $^7$Be flux could be between 2% and 97% of the standard value.

• The "luminosity constraint" should take account of the fact that there must be a pp (or pep) neutrino for each $^7$Be (or $^8$B ) neutrino, if the solar luminosity is currently being supplied by nuclear fusion reactions among light elements. Eqs. (7)–(10), when taken together, constitute a statement of the luminosity constraint that is estimated to be accurate to better than 1%. At 95% confidence level, the upper and lower bounds on the allowed pp neutrino creation rate are determined by the luminosity constraint and the allowed upper limit on the $^7$Be neutrino creation rate is established by the luminosity constraint.

• For pp and $^7$Be neutrinos, the luminosity constraint, not the four operating solar neutrino experiments, sets the most stringent limits on the allowed range of creation rates in the solar interior. The ratio of the creation rate in the center of the Sun to the rate predicted in the 1995 standard solar model varies between 0.55 to 1.08 for pp neutrinos and between 0.0 to 6.35 for the $^7$Be neutrinos assuming the CNO neutrino flux doesn't change from its predicted value in the standard solar model.

• For $^8$B neutrinos, the operating experiments establish the allowed range of neutrino creation rates in the solar interior. For MSW oscillations, the ratio of the $^8$B neutrino creation rate in the interior of the Sun to the rate predicted in the 1995 standard solar model varies between 0.4 to 2.4. For vacuum oscillations, the ratio of the allowed flux to the flux predicted in the 1995 solar model varies between 0.6 to 2.8 for the $^8$B neutrinos.

• The predicted event rates for the SuperKamiokande, SNO, BOREXINO, ICARUS, HELLAZ, and HERON experiments are given in Table VII for the best-fit small-angle MSW, large-angle MSW, and vacuum neutrino oscillation solutions. The best-estimates for SuperKamiokande, SNO, and ICARUS do not depend sensitively upon the assumed oscillation scenario. However, the range of expected event rates is rather large in all of the experiments if we allow the neutrino parameters to vary over the predicted 95% C.L. .

• Futuristic calculations shown in Fig. 5–Fig. 8 illustrate some of the power of the planned experiments. For example, a 5% measurement of the total event rate with SNO will, if the small-angle MSW solution is correct, eliminate almost the entire large-angle MSW solution (see Figure 5b). BOREXINO will eliminate the large mixing angle (Figure 5c) and the vacuum oscillation solutions (see Figure 6c). If the vacuum oscillation parameters that best fit the currently-operating experiments are the correct solution of the solar neutrino problems, a 5% measurement of the total event rate in SNO or BOREXINO could essentially eliminate the small angle MSW solution (see Figure 7). Moreover, a 5% measurement with SuperKamiokande, SNO, or BOREXINO will also eliminate most of the existing neutrino parameter space for vacuum solutions (see Fig. 8). There is a significant improvement in the calculated ability of future experiments to discriminate between different neutrino mixing parameters when the assumed measurement errors are reduced from 10% to 5%, but there is very little additional improvement when the errors are reduced from 5% to 1%.

- The double ratio of neutral current to charged current rates in the SNO detector is, by definition [Eq. (16)], unity if no oscillations occur and has a 95% confidence range of $3.1^{+1.8}_{-1.3}$ for the small mixing angle MSW solution, $4.4^{+2.0}_{-1.4}$ for the large mixing angle MSW solution, and $5.2^{+5.6}_{-2.9}$ for the vacuum oscillation solutions. These results are relatively insensitive to changes in the solar model and suggest that an accurate measurement of the ratio of the total neutral current to charged current event rates may well be capable of distinguishing between no oscillation solutions and oscillation solutions (either MSW or vacuum).

## ACKNOWLEDGMENTS

J.N.B. acknowledges support from NSF grant #PHY92-45317. The work of P.I.K. was partially supported by funds from the Institute for Advanced Study. We thank S. Glashow for asking a stimulating question that sparked our investigation of future detectors, K.S. Babu, M. Fukugita, C. Kolda, E. Lisi, and S. Parke for valuable comments and very helpful suggestions on the original manuscript, and A. Gould and W. Press for important discussions of statistical questions.

TABLES

TABLE I. Solar neutrino data used in the analysis. The experimental results are given in SNU for all of the experiments except Kamiokande, for which the result is expressed the measured $^8$B flux above 5 MeV in units of cm$^{-2}$s$^{-1}$ at the earth. The ratios of the measured values to the corresponding predictions in the standard solar model of ref. [5] are also given. The result cited for the Kamiokande experiment assumes that the shape of the $^8$B neutrino spectrum is not affected by physics beyond the standard electroweak model.

| Experiment | Result (1$\sigma$) | Exp. Result/Th. Calculation | Reference |
|---|---|---|---|
| HOMESTAKE | $2.55 \pm 0.17$(stat) $\pm 0.18$(syst) | $0.27 \pm 0.03$ | [1] |
| GALLEX | $77.1 \pm 8.5$(stat) $\pm^{+4.4}_{-5.4}$ (syst) | $0.56 \pm 0.07$ | [2] |
| SAGE | $69 \pm 11$(stat)$^{+5}_{-7}$(syst) | $0.50 \pm 0.09$ | [3] |
| KAMIOKANDE | $[2.89^{+0.22}_{-0.21}$ (stat) $\pm 0.35$ (syst)$] \times 10^6$ | $0.44 \pm 0.06$ | [4] |

TABLE II. The Standard Model Solar Neutrino Fluxes. The neutrino fluxes given in this table are from the 1995 standard solar model of Bahcall and Pinsonneault [5], which includes both helium and heavy element diffusion. We also use in constructing Figures 1 and 2 a 1988 [13] and a 1992 [27] standard solar model.

| Source | Energy (MeV) | Flux ($10^{10}$ cm$^{-2}$s$^{-1}$) |
|---|---|---|
| pp | $\leq 0.42$ | $5.91(1.00^{+0.01}_{-0.01})$ |
| pep | $1.44$ | $1.40 \times 10^{-2}(1.00^{+0.01}_{-0.02})$ |
| $^7$Be | $0.86, 0.38$ | $5.15 \times 10^{-1}(1.00^{+0.06}_{-0.07})$ |
| $^8$B | $\lesssim 15$ | $6.62 \times 10^{-4}(1.00^{+0.14}_{-0.17})$ |
| $^{13}$N | $\leq 1.2$ | $6.18 \times 10^{-2}(1.00^{+0.17}_{-0.20})$ |
| $^{15}$O | $\leq 1.7$ | $5.45 \times 10^{-2}(1.00^{+0.19}_{-0.22})$ |
| $^{17}$F | $\leq 1.7$ | $6.48 \times 10^{-4}(1.00^{+0.15}_{-0.19})$ |

TABLE III. MSW limiting $\nu_e$ fluxes. The minimum and maximum values (at 95% C.L.) are given for the ratios, R, of the measured to the predicted (1995 solar model) event rates in the four operating solar neutrino experiments.

| Experiment/SSM | R(pp)$_{min}$ | R(pp)$_{max}$ | R($^7$Be)$_{min}$ | R($^7$Be)$_{max}$ | R($^8$B)$_{min}$ | R($^8$B)$_{max}$ |
|---|---|---|---|---|---|---|
| HOMESTAKE | – | – | 0.005 | 0.65 | 0.16 | 0.56 |
| SAGE/GALLEX | 0.50 | 0.997 | 0.01 | 0.65 | 0.16 | 0.54 |
| KAMIOKANDE | – | – | – | – | 0.28 | 0.64 |
| SUPERKAMIOKANDE | – | – | – | – | 0.28 | 0.60 |
| SNO | – | – | – | – | 0.16 | 0.55 |
| BOREXINO | – | – | 0.21 | 0.72 | – | – |
| ICARUS | – | – | – | – | 0.16 | 0.57 |
| HELLAZ/HERON | 0.65 | 0.998 | – | – | – | – |

TABLE IV. Vacuum oscillations limiting $\nu_e$ fluxes. The minimum and maximum values (at 95% C.L.) are given for the ratios, R, of the measured to the predicted (1995 solar model) event rates in the four operating solar neutrino experiments.

| Experiment/SSM | R(pp)$_{min}$ | R(pp)$_{max}$ | R($^7$Be)$_{min}$ | R($^7$Be)$_{max}$ | R($^8$B)$_{min}$ | R($^8$B)$_{max}$ |
|---|---|---|---|---|---|---|
| HOMESTAKE | – | – | 0.083 | 0.98 | 0.12 | 0.43 |
| GALLEX/SAGE | 0.49 | 0.67 | 0.11 | 0.97 | 0.13 | 0.43 |
| KAMIOKANDE | – | – | – | – | 0.24 | 0.53 |
| SUPERKAMIOKANDE | – | – | – | – | 0.25 | 0.56 |
| SNO | – | – | – | – | 0.09 | 0.42 |
| BOREXINO | – | – | 0.29 | 0.98 | – | – |
| ICARUS | – | – | – | – | 0.11 | 0.43 |
| HELLAZ/HERON | 0.64 | 0.77 | – | – | – | – |

TABLE V. Detector independent limiting $\nu_e$ fluxes. The limiting fractional fluxes (relative to the SSM) are given, independent of detector sensitivities [see Eq. (2)] for both MSW and vacuum neutrino solutions.

|        | R(pp)$_{min}$ | R(pp)$_{max}$ | R($^7$Be)$_{min}$ | R($^7$Be)$_{max}$ | R($^8$B)$_{min}$ | R($^8$B)$_{max}$ |
|--------|---------------|---------------|-------------------|-------------------|------------------|------------------|
| MSW    | 0.52          | 0.997         | 0.02              | 0.65              | 0.11             | 0.40             |
| vacuum | 0.49          | 0.67          | 0.13              | 0.97              | 0.27             | 0.60             |

TABLE VI. Coefficients for Luminosity Constraint. The coefficients are given in MeV.

| Flux       | $\alpha$ | Flux      | $\alpha$ |
|------------|----------|-----------|----------|
| pp         | 13.097   | $^{13}$N  | 3.457    |
| pep        | 11.918   | $^{15}$O  | 21.572   |
| $e^- + {}^7$Be | 12.500 | $^{17}$F  | 2.363    |
| $^8$B      | 6.655    | hep       | 10.170   |

TABLE VII. Predicted event rates in future solar neutrino experiments expressed as ratios to the event rates expected from the standard solar model [5]. The values of $\Delta m^2$ and $\sin^2 2\theta$ used to minimize $\chi^2$ for the existing four experiments are given in Eqs. 1–3.

|                | MSW (Small Mixing)        | MSW (Large Mixing)        | Vacuum Oscillations       |
|----------------|---------------------------|---------------------------|---------------------------|
| SUPERKAMIOKANDE | $0.41^{+0.19}_{-0.13}$   | $0.34^{+0.09}_{-0.06}$   | $0.31^{+0.25}_{-0.06}$   |
| SNO            | $0.32^{+0.23}_{-0.16}$   | $0.22^{+0.23}_{-0.06}$   | $0.19^{+0.23}_{-0.10}$   |
| BOREXINO       | $0.22^{+0.50}_{-0.01}$   | $0.59^{+0.13}_{-0.14}$   | $0.69^{+0.29}_{-0.40}$   |
| ICARUS         | $0.34^{+0.23}_{-0.18}$   | $0.22^{+0.11}_{-0.06}$   | $0.23^{+0.20}_{-0.12}$   |
| HELLAZ/HERON   | $0.96^{+0.04}_{-0.31}$   | $0.73^{+0.06}_{-0.08}$   | $0.67^{+0.10}_{-0.03}$   |

TABLE VIII. The ratio of neutral to charged current event rates in the SNO detector for different solutions of the solar neutrino problem. The ranges of the ratio (Min and Max) have been determined by varying $\delta m^2$ and $\sin^2 2\theta$ within the 95% allowed regions for each solution. The electron energy threshold was assumed to be either 5 MeV (upper line) or 6 MeV (lower line).

| Scenario | (NC/CC) | (NC/CC)$_{\rm DR}$ | Min(NC/CC)$_{\rm DR}$ | Max(NC/CC)$_{\rm DR}$ |
|---|---|---|---|---|
| Standard model | 0.44 | 1.00 | 1.00 | 1.00 |
|  | 0.49 | 1.00 | 1.00 | 1.00 |
| Small angle MSW | 1.3 | 3.1 | 1.8 | 4.9 |
|  | 1.5 | 3.0 | 1.8 | 4.7 |
| Large angle MSW | 1.9 | 4.4 | 3.0 | 6.4 |
|  | 2.2 | 4.5 | 3.1 | 6.5 |
| Vacuum oscillations | 2.3 | 5.2 | 2.3 | 11 |
|  | 2.3 | 4.8 | 2.4 | 11 |

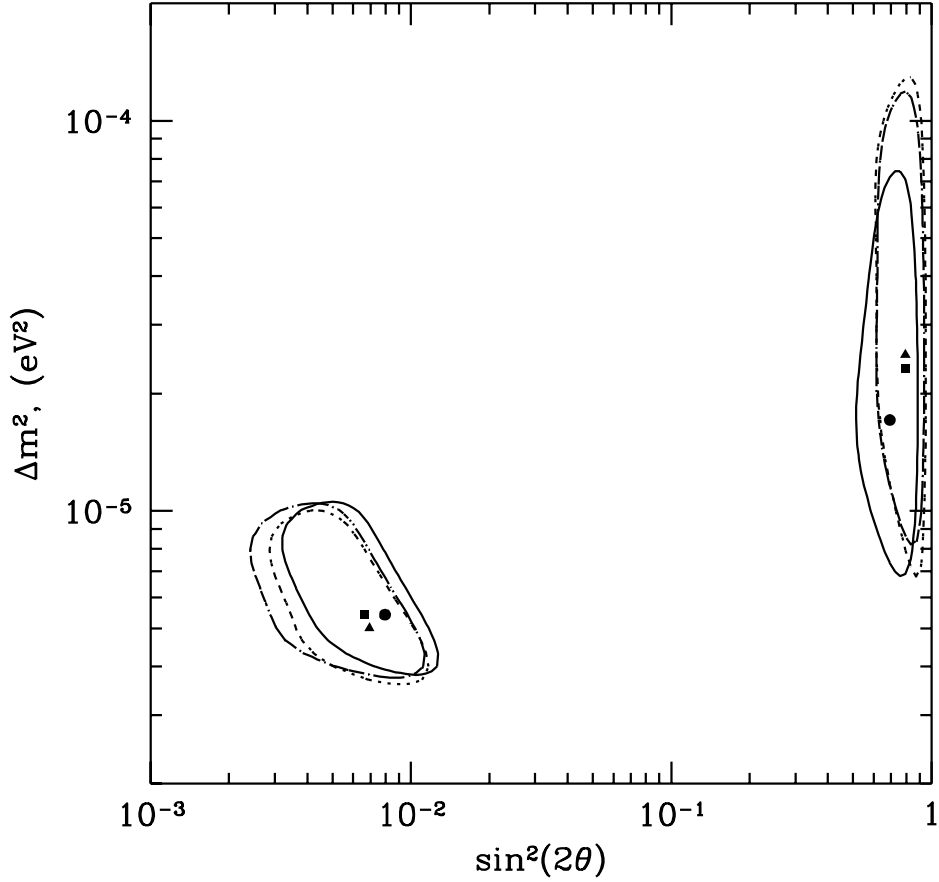

FIG. 1. Allowed MSW Solutions. The allowed regions at 95% C.L. are shown for $\sin^2 2\theta$ and $\Delta m^2$ with the MSW solution of the solar neutrino problems. The dashed and dash-dotted line contours are for the solar models [13] (1988) and [27] (1992); the full line contour is for the most recent solar model [5] (1995). The points where $\chi^2$ has a local minimum are indicated by a triangle (1988 solar model), a square (1992 model), and a circle (1995 model).

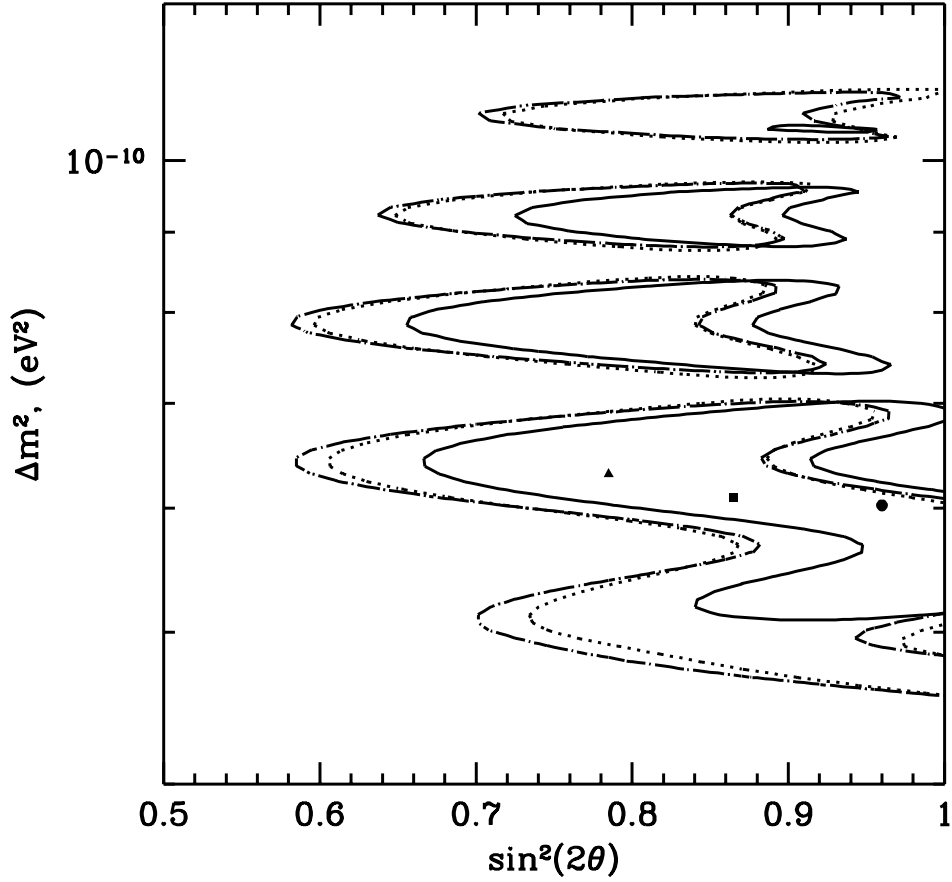

FIG. 2. Allowed Vacuum Solutions. The allowed regions at 95% C.L. are shown for the vacuum neutrino oscillation solution of the solar neutrino problems. The symbols have the same meaning as for Fig. 1.

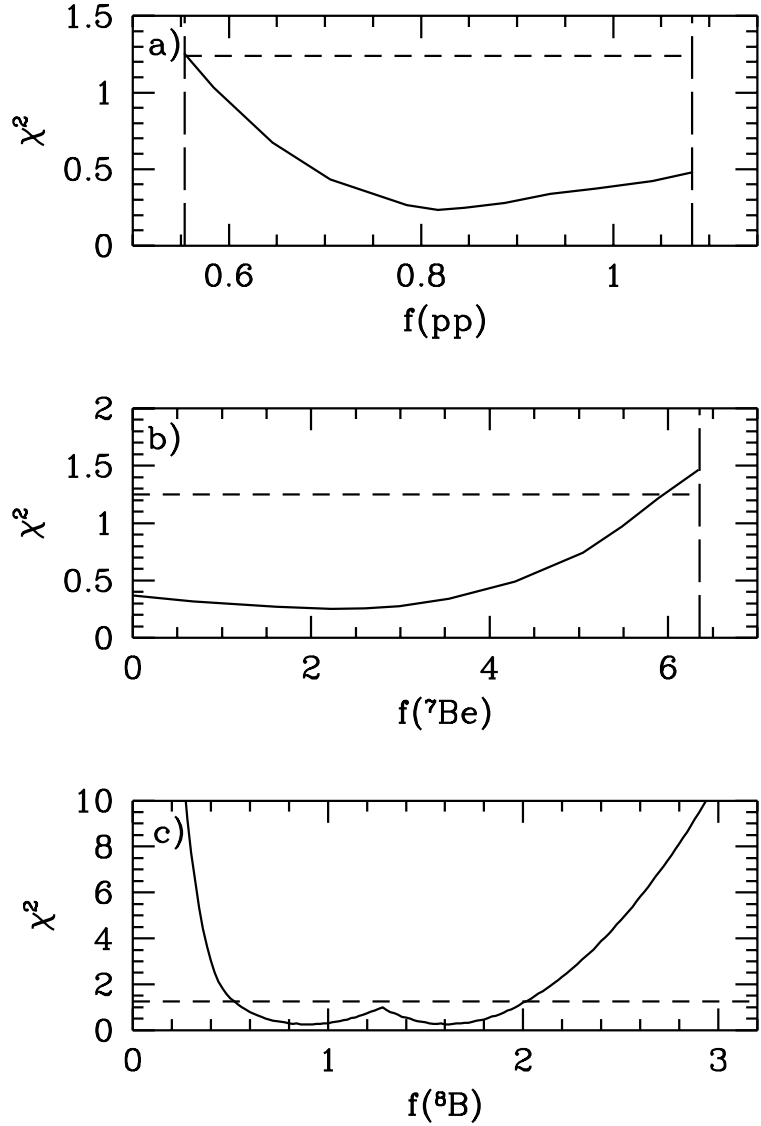

FIG. 3. The minimum $\chi^2$ as a function of one neutrino flux for the MSW solution. One neutrino flux and the neutrino oscillation parameters, $\sin^2 2\theta$ and $\Delta m^2$, are treated as free parameters. In Fig. 3a, the pp neutrino flux is treated as a free parameter; in Fig. 3b and Fig. 3c, the $^7$Be and $^8$B fluxes are treated as free parameters. The horizontal line marks $\chi^2 = \chi^2_{\min} + 1.0$. The vertical lines represent the luminosity constraint, Eqs. (7)–(10).

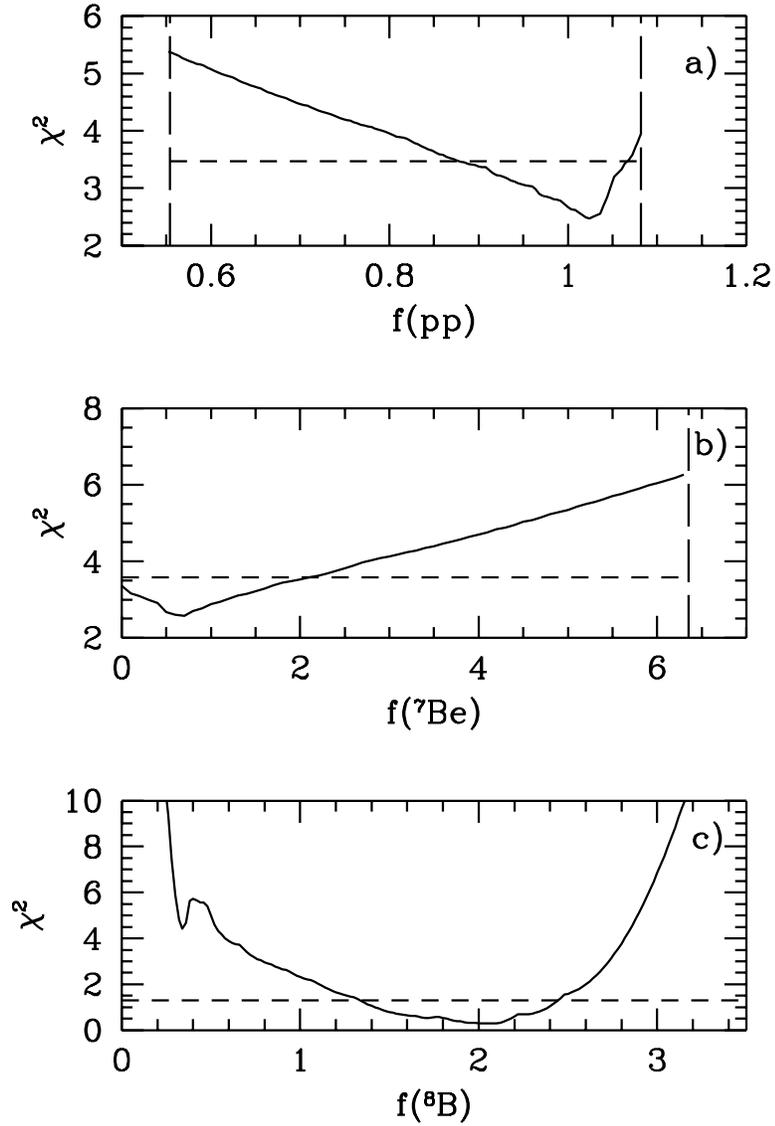

FIG. 4. The minimum $\chi^2$ as a function of one neutrino flux for the vacuum oscillation solution. One neutrino flux and the neutrino oscillation parameters, $\sin^2 2\theta$ and $\Delta m^2$, are treated as free parameters. In Fig. 4a, the pp neutrino flux is treated as a free parameter; in Fig. 4b and Fig. 4c, the $^7$Be and $^8$B fluxes are treated as free parameters. The horizontal line marks $\chi^2 = \chi^2_{\min} + 1.0$. The vertical lines represent the luminosity constraint, Eqs. (7)–(10).

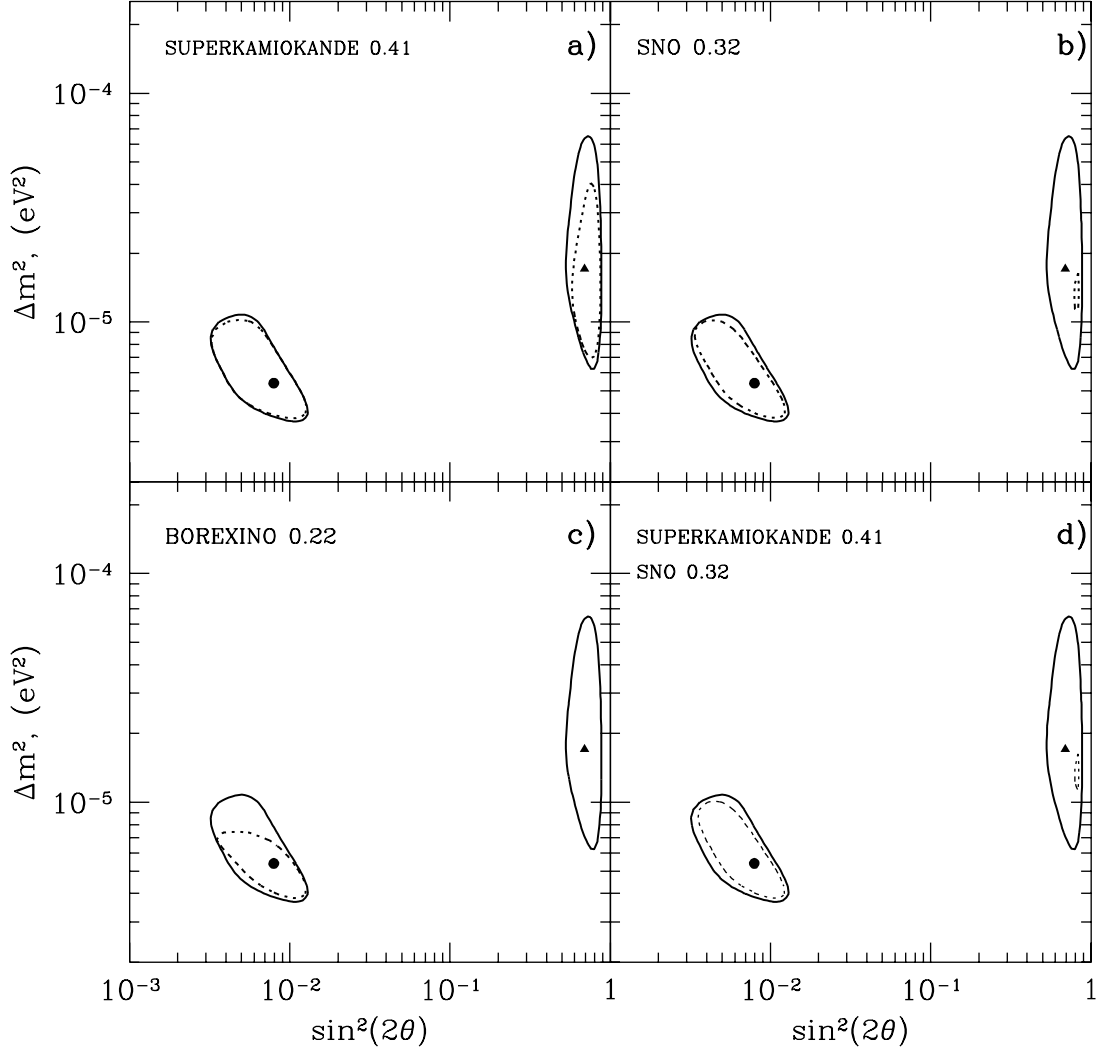

FIG. 5. Future Experiments: The Small Mixing Angle MSW Solution Assumed Correct and the 'Future' Allowed MSW Solutions. Assuming the correctness of the small mixing angle MSW solution, the results of future solar neutrino experiments are calculated. A 5% error, normally-distributed, is assumed for each of the future experiments. The 95% C.L. allowed regions for $\Delta m^2$ and $\sin^2 2\theta$ are calculated in a)-c) for four operating and one new experiment and in d) for four operating and two new experiments. The regions enclosed within the dark lines are permitted by the four operating experiments and the regions enclosed within the dashed lines are permitted by the combined operating and future experiments.

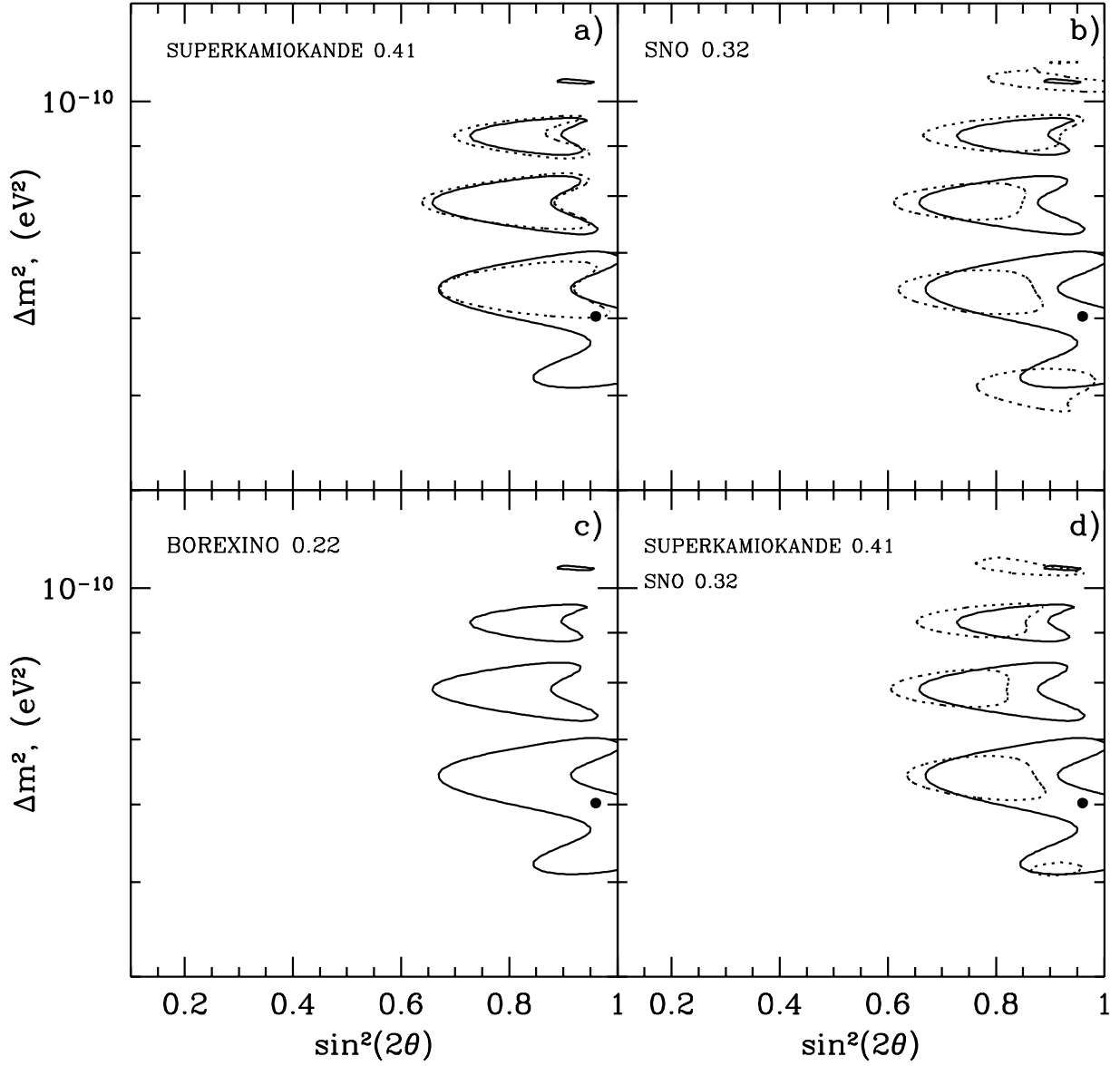

FIG. 6. Future Experiments: The Small Mixing Angle MSW Solution Assumed Correct and the 'Future' Allowed Vacuum Oscillation Solutions. The caption for this figure is the same as for Figure 5 except that the allowed regions for vacuum neutrino oscillations are calculated for the combined operating and future experiments.

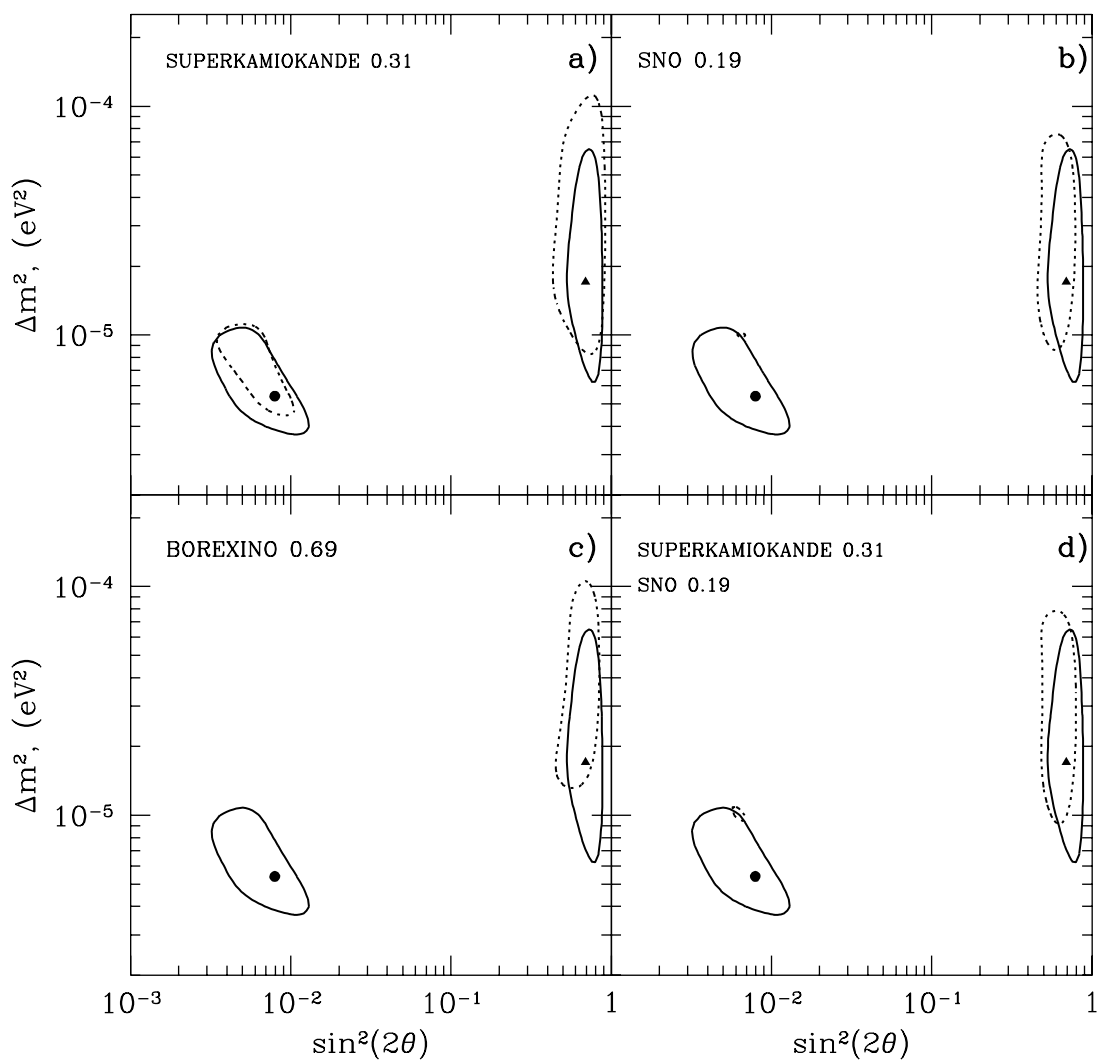

FIG. 7. Future Experiments: Vacuum Mixing Assumed Correct and the 'Future' Allowed MSW Solutions. Assuming the correctness of the minimum $\chi^2$ vacuum neutrino solution, the results of future solar neutrino experiments are calculated. A 5% error, normally-distributed, is assumed for each of the future experiments. The 95% C.L. allowed regions for $\Delta m^2$ and $\sin^2 2\theta$ are calculated in a)-c) for four operating and one new experiment and in d) for four operating and two new experiments. The regions enclosed within the dark lines are permitted by the four operating experiments and the regions enclosed within the dashed lines are permitted by the combined operating and future experiments.

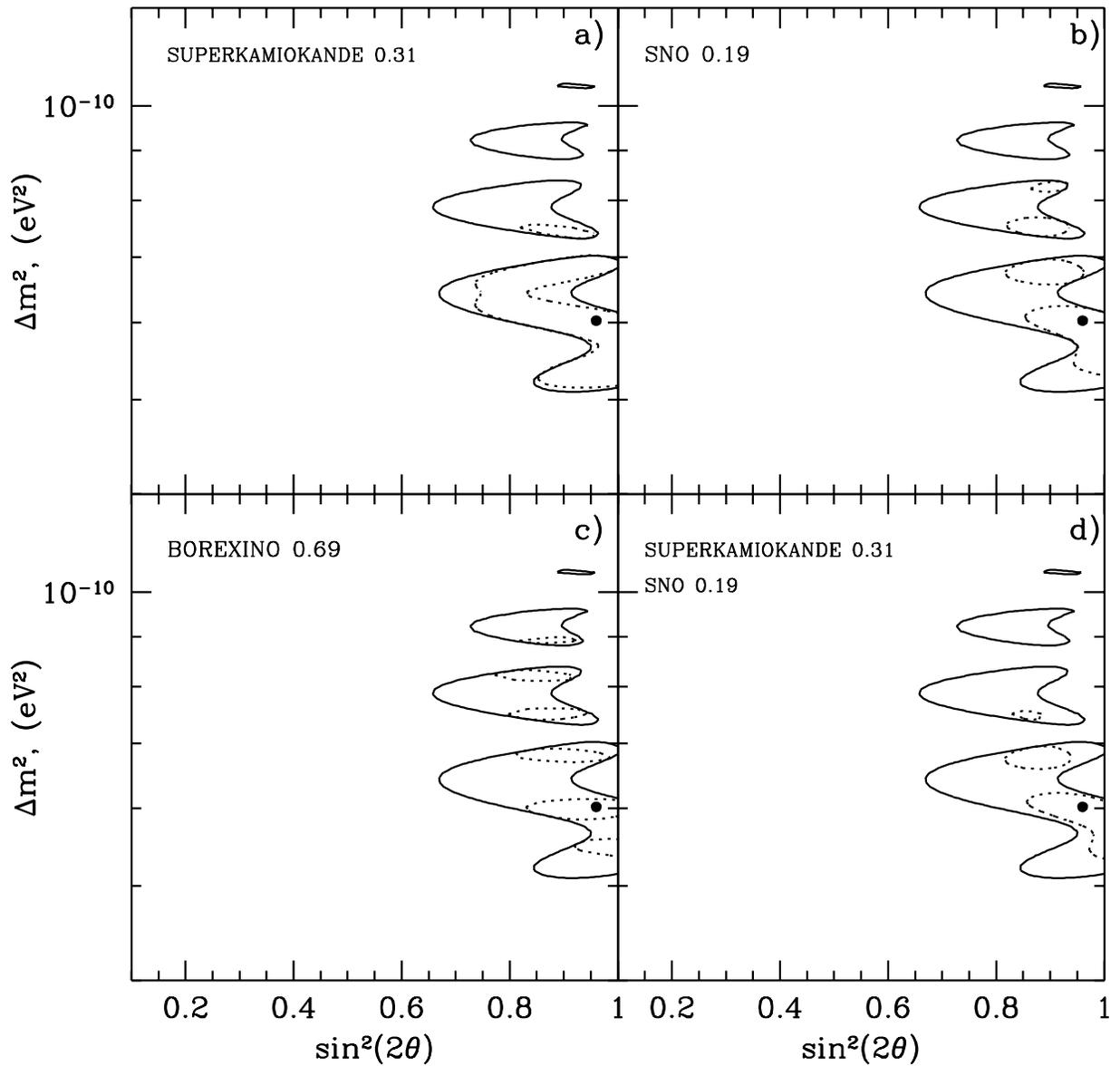

FIG. 8. Future Experiments: Vacuum Mixing Assumed Correct and the 'Future' Allowed Vacuum Oscillation Solutions. The caption for this figure is the same as for Figure 7 except that the allowed regions for vacuum neutrino oscillations are calculated for the combined operating and future experiments.